\documentclass[a4paper,fleqn,usenatbib]{mnras}

\usepackage{times,txfonts}
\usepackage[T1]{fontenc}
\usepackage{ae,aecompl}

\usepackage{graphicx}	

\def\HI{{\rm H{\small I }}}

\title[21~cm Power Spectrum Upper Limits from Quasars]{Upper Limits on the 21~cm Power Spectrum at $z=5.9$ from Quasar Absorption Line Spectroscopy}

\author[J. C. Pober et al.]{
Jonathan C. Pober,$^{1}$\thanks{E-mail: Jonathan\_Pober@brown.edu}
Bradley Greig,$^{2}$
and Andrei Mesinger$^{2}$
\\
$^{1}$Department of Physics, Brown University, Providence, RI 02912, USA\\
$^{2}$Scuola Normale Superiore, Piazza dei Cavalieri 7, I-56126 Pisa, Italy
}

\date{Accepted 2016 August 1. Received 2016 July 11; in original form 2016 10 May}

\pubyear{2016}

\begin{document}
\label{firstpage}
\pagerange{\pageref{firstpage}--\pageref{lastpage}}
\maketitle

\begin{abstract}
We present upper limits on the 21~cm power spectrum at $z=5.9$ calculated from the model-independent
limit on the neutral fraction of the 
intergalactic medium of $x_{\HI} < 0.06 + 0.05\ (1\sigma)$ derived from dark pixel 
statistics of quasar absorption spectra.  Using \texttt{21CMMC}, a Markov chain Monte Carlo Epoch of
Reionization analysis code, we explore
the probability distribution of 21~cm power spectra consistent with this constraint on the neutral
fraction.  We present 99 per cent confidence upper limits of $\Delta^2(k) < 10$ to $20\ {\rm mK}^2$
over a range of $k$ from 0.5 to $2.0\ h{\rm Mpc}^{-1}$,
with the exact limit dependent on the sampled $k$ mode.
This limit can be used as a null test for 21~cm experiments: a detection of power at $z=5.9$ in excess
of this value is highly suggestive of residual foreground contamination or other systematic errors
affecting the analysis.
\end{abstract}



\section{Introduction}

Observations of the highly redshifted 21~cm line of neutral hydrogen have become recognized as 
one of the most promising probes for exploring the early universe.  A particular focus
of early experiments is to detect the spatial power spectrum of this 
emission from the Epoch of Reionization (EoR), where the contrast between neutral and ionized regions
of the intergalactic medium (IGM) can produce a relatively strong signal.
Several experiments have conducted lengthy observation campaigns with the goal of making a first
detection, including 
the Giant Metre-Wave Radio Telescope (GMRT; \citealt{swarup_et_al_1991})\footnote{http://gmrt.ncra.tifr.res.in/},
the LOw Frequency ARray (LOFAR; \citealt{van_haarlem_et_al_2013})\footnote{http://www.lofar.org/}, 
the Murchison Widefield Array (MWA; \citealt{tingay_et_al_2013a})\footnote{http://www.mwatelescope.org/}, 
and the Donald C. Backer Precision Array for Probing the Epoch of Reionization
(PAPER; \citealt{parsons_et_al_2010})\footnote{http://eor.berkeley.edu/}.
To date, only upper limits have been placed on the cosmological signal, with the best constraints
coming at $z=8.4$ from PAPER \citep{ali_et_al_2015}.  While not sensitive enough to detect
the most generic reionization signals, these measurements were able to place lower limits on the IGM
temperature of $\sim 10$~K, beginning to rule out some models of ``cold reionization" 
\citep{pober_et_al_2015,greig_et_al_2016}.

As these experiments continue to collect data and improve analysis techniques, the possibility
of a first detection of the 21~cm signal seems near.  A first detection, however, will likely
be of low to moderate significance and limited to a narrow range of redshifts and Fourier $k$ modes.
And, nearly all of the upper limits published to date have seen systematic biases that are not
consistent with noise \citep{parsons_et_al_2014,jacobs_et_al_2015,ali_et_al_2015,dillon_et_al_2015b,trott_et_al_2016}.
It is therefore exceedingly germane to ask the question: how can one distinguish an early, 
low-significance 21~cm detection from systematics caused by foregrounds or other instrumental
effects?

There are several features intrinsic to the cosmological 21~cm signal that could 
be used to distinguish it from foregrounds.
One of the most commonly proposed is to observe the ``rise and fall" of the 21~cm signal
with frequency.  The hydrogen signal is predicted to peak when the ionization
fraction of the IGM is 50 per cent \citep{lidz_et_al_2008,bittner_and_loeb_2011} or somewhat
thereafter depending on the relative efficiency of X-ray heating
and UV reionization \citep{mesinger_et_al_2013,mesinger_et_al_2016}; therefore, a
cosmological signal should show a peak in the power spectrum as a function 
of redshift (and therefore a peak as a function of frequency,
since observed frequency corresponds to redshift of the 21~cm line).  
Foregrounds on the other hand are dominated by synchrotron emission,
which smoothly increases in power towards lower frequencies.\footnote{While many sources do show
a turnover in their synchrotron spectra due to self-absorption, this typically happens at much lower
frequencies than the $100-200$~MHz range where one might expect to see a reionization signal.}
However, prospects for detecting this peak are perhaps more limited than initially expected: due to the
steep increase of sky noise towards lower frequencies ($T_{\rm sky} \propto \nu^{-2.55}$), the 
uncertainties on higher redshift 21~cm measurements grow rapidly, limiting the significance with
which an actual rise-and-fall can be detected \citep{pober_et_al_2014}.
While a decline in signal strength towards lower frequencies can be detected with high significance,
the same spectral behavior is shared with foregrounds; therefore, the peak in frequency is the
clear discriminant.
Moreover, if inhomogeneous recombinations play an import role in the reionization process, 
they can decrease the power on large scales and reduce the prominence of the peak versus redshift,
resulting in a redshift evolution of the cosmological signal that can be more modest than initially predicted,
\citep{sobacchi_and_mesinger_2014}.

Another potential discriminator could be the (nearly) spherically symmetric shape of the 21~cm signal
in $k$ space, since foregrounds have drastically different angular and spectral properties.  However,
first generation experiments do not have the sensitivity to measure high-$k$ transverse modes
of the power spectrum, so this spherical symmetry will be very difficult to detect \citep{pober_2015}.
 The shape of the spherically averaged 1D 
21~cm power spectrum is itself relatively featureless, with exception
of a ``knee" on very large scales. Foregrounds are most problematic on these scales,
although it may be possible to recover this feature in some cases \citep{greig_and_mesinger_2015}.
Its usefulness as a robust indicator of a cosmological signal may be limited for first-generation
experiments.

Cross-correlation studies present another promising tool for confirming the presence of cosmological
emission in 21~cm data.  If the signal is truly of cosmological origin, then it should spatially
correlate with other tracers of large scale structure, including galaxies and Lyman-$\alpha$ emitters
\citep{furlanetto_and_lidz_2007,sobacchi_et_al_2016,vrbanec_et_al_2016}, the near infrared background
\citep{wyithe_et_al_2007,fernandez_et_al_2014,mao_2014}, and ``intensity maps" of
other spectral lines such as CO and CII
\citep{gong_et_al_2011,lidz_et_al_2011,gong_et_al_2012}.  However, these correlation studies either
require larger volume surveys than are possible with existing instruments or new, dedicated
experiments in the case of intensity mapping.  

The goal of this letter is to propose and quantify a potential test for the robustness of a
purported initial 21~cm detection.  It has been recognized that observations at redshifts \emph{after} 
the end of reionization might provide a null result.  It is the contrast between the neutral regions
and ionized bubbles within the IGM that leads to large spatial fluctuations in the hydrogen signal,
and therefore a relatively bright power spectrum during reionization.  Once the IGM is completely
ionized, this contrast disappears, and hence the signal becomes undetectable.\footnote{While neutral
hydrogen still remains in shelf-shielded structures (i.e. galaxies), this signal is orders of magnitude
fainter than the signal during reionization 
\citep{barkana_and_loeb_2007,crociani_et_al_2011,ansari_et_al_2012b,sobacchi_and_mesinger_2014}.
The reason this signal could be detectable at lower redshifts (e.g. $z\sim1$) is because of the significant
decrease in the brightness of foregrounds and the system noise at higher frequencies 
\citep{pober_et_al_2013a}.}
In this work, we turn this expectation into a quantitative prediction for the maximum possible
brightness of the 21~cm power spectrum at $z=5.9$.  This particular redshift is especially useful
for these studies, as it falls within the range accessible to current 21~cm experiments,
but also has significant observations at other wavelengths that allow for a 
non-trivial constraint on the 21~cm power.  To perform this calculation, we use the model-independent
constraint on the neutral fraction at this redshift from \cite{mcgreer_et_al_2015}.
By measuring the fraction of dark (i.e. zero flux) pixels in the Lyman-$\alpha$ and Lyman-$\beta$ forests in 
the spectra of high-$z$ quasars, they place a limit on the neutral fraction
$x_{\HI} < 0.06 + 0.05\ (1\sigma)$ at $z=5.9$.  This technique does not make assumptions about
whether dark pixels are caused by genuine cosmic \HI regions from an incomplete reionization,
or from residual neutral hydrogen in the ionized IGM (note that trace amounts of {{\rm H{\small I}}}, at the level
of $10^{-4}$, are sufficient to saturate the Lyman-$\alpha$ forest).
Hence, it is a conservative limit, but avoids model-driven assumptions.  To turn this
constraint into a limit on the 21~cm power spectrum, we use the \texttt{21CMMC} 
\citep{greig_and_mesinger_2015}\footnote{https://github.com/BradGreig/21CMMC} EoR analysis code to perform
a Markov chain Monte Carlo (MCMC) exploration of possible 21~cm power spectra given this neutral fraction limit.

The structure of this letter is as follows.  In \S\ref{sec:21cmmc}, we describe the modeling
of the 21~cm power spectrum performed by \texttt{21CMMC}.  In \S\ref{sec:results}, we present
the resultant power spectrum limits.  We discuss the implications of this result in \S\ref{sec:discussion}
and conclude in \S\ref{sec:conclusions}.  Unless otherwise stated, 
we assume a $\Lambda$CDM cosmological model with the Planck 2015
cosmlogical parameter values:
$h = 0.6781$, $\Omega_M = 0.308$, $\Omega_b = 0.0484$, $\Omega_{DE} = 0.692$, and $\sigma_8 = 0.8149$
\citep{planck_2015_XIII}.  

\section{Power Spectrum Modeling}
\label{sec:21cmmc}

\texttt{21CMMC} is an MCMC sampler of 
\texttt{21cmFAST}\footnote{http://homepage.sns.it/mesinger/DexM\_\_\_21cmFAST.html}, a
semi-numerical simulation code for quickly generating realizations of the ionization history
of the universe and associated 21~cm signal.  We highlight the relevant details of each code here,
but refer the reader to \cite{greig_and_mesinger_2015} for more details about \texttt{21CMMC}
and to \cite{mesinger_and_furlanetto_2007} and \cite{mesinger_et_al_2011} for a complete
description of \texttt{21cmFAST}. 

\texttt{21cmFAST} generates IGM density, velocity, source, and ionization fields by creating a
realization of the 3D linear density field within a cubic volume and evolving it following
the Zel'dovich approximation \citep{zeldovich_1970}.
The ionization field is estimated by applying the excursion set approach of \cite{furlanetto_et_al_2004},
which compares the time-integrated number of ionizing photons to the number of baryons within 
regions of decreasing radius, to perturbed 3D density fields.
These formalisms and approximations allow for rapid generation of simulated 21~cm cubes
(of order 45~seconds for a single redshift realization).  When compared with more rigorous
simulations, \texttt{21cmFAST} has been shown to produce power spectra accurate to within
10 to 30 per cent on the scales of interest for reionization.  The dominant uncertainties
come from the ionization field; the underlying matter power spectra agree with N body
simulations to better than the per cent level \citep{mesinger_et_al_2011}.
  
The primary physics of reionization are described in \texttt{21cmFAST}
with three key parameters: $\zeta$, an ionizing efficiency which converts baryonic mass into a number
of ionizing photons; $T_{\rm vir,min}$, a minimum virial temperature below which halos do
not produce ionizing photons; and $R_{\rm mfp}$, the maximum horizon of ionizing photons through
ionized regions of the IGM.\footnote{In the literature on \texttt{21cmFAST}, this parameter is often referred to as a mean
free path for ionizing photons, but in actuality, it sets a sharp cutoff for the maximum distance an ionizing
photon can travel before it is absorbed.}
Although this three parameter model is an oversimplification of reionization physics,
it provides a physically-motivated basis set for sampling 21~cm power spectra during the EoR,
which is the primary requirement for this work.  Moreover, the resulting 21~cm power spectra from
this simple model agree with those from complex semi-analytical models when using a comparable
average halo mass of reionization galaxies \citep{geil_et_al_2015}.
In this simulation, we use a box of 500~Mpc on a side with $256^3$ voxels.
Note that we do not calculate spin temperature fluctuations in our \texttt{21cmFAST} simulations, as
the spin temperature of the IGM is very likely well above the Cosmic Microwave Background (CMB) temperature
at $z=5.9$;\footnote{While there are no direct measurements of the temperature of putative cosmic \HI
regions at this redshift, theoretical expectations are that X-ray heating from collapsed stellar remnants 
\citep{furlanetto_2006,mcquinn_2012,mesinger_et_al_2013} brings the spin temperature of the gas
well above that of the CMB.  Even in the absence of this mechanism,
shock heating should be enough to raise the spin temperature above the CMB
temperature at this low redshift \citep{mcquinn_and_oleary_2012}.}
in this limit, the 21~cm signal is independent of the spin temperature.

\texttt{21CMMC} uses MCMC sampling to explore the likelihood for
the three EoR parameters $\zeta,\ T_{\rm vir,min}$, and $R_{\rm mfp}$
(generating a \texttt{21mFAST} realization for each set) given
power spectrum measurements from a 21~cm experiment.  Here, we use a modified version of the
code, as we do not have power spectrum measurements to compare with.  Rather, we explore combinations
of EoR parameters that recover neutral fractions consistent with the \cite{mcgreer_et_al_2015} limits
at $z = 5.9$.  This limit is not enough to place significant constraints on any of the three EoR
parameters (Greig and Mesinger, in prep.); 
however, it does place significant constraints on the maximum allowable amplitude of
the 21~cm power spectrum.  It is this limit that we explore in this work. 

\section{Results}
\label{sec:results}

Using \texttt{21CMMC}, we explore the likelihood function for the amplitude of the 21~cm power
spectrum using the \cite{mcgreer_et_al_2015} limit and uniform priors on the EoR
parameters over the ranges $5 \leq \zeta \leq 200$, $10^4\ {\rm K} \leq T_{\rm vir,min} \leq 10^{5.7}\ {\rm K}$,
and $5\ {\rm Mpc} \leq R_{\rm mfp} \leq 40\ {\rm Mpc}$. See \cite{mesinger_et_al_2012}, \cite{pober_et_al_2014}, 
\cite{greig_and_mesinger_2015}, and references therein for the rationale behind these ranges.
Note that the $R_{\rm mfp}$ parameter roughly 
corresponds to a time-averaged mean free path through the ionized IGM, and so is likely to be at or 
below the measurements at $z \sim 6$ from, e.g., \cite{songaila_and_cowie_2010}.  
We also note that the evolution and structure of reionization in the simulations
is insensitive to this choice beyond $R_{\rm mfp}\gtrsim30$ Mpc
(see, e.g., \citealt{furlanetto_and_mesinger_2009} and \citealt{alvarez_and_abel_2012}).

\begin{figure}
\includegraphics[width=3.5in]{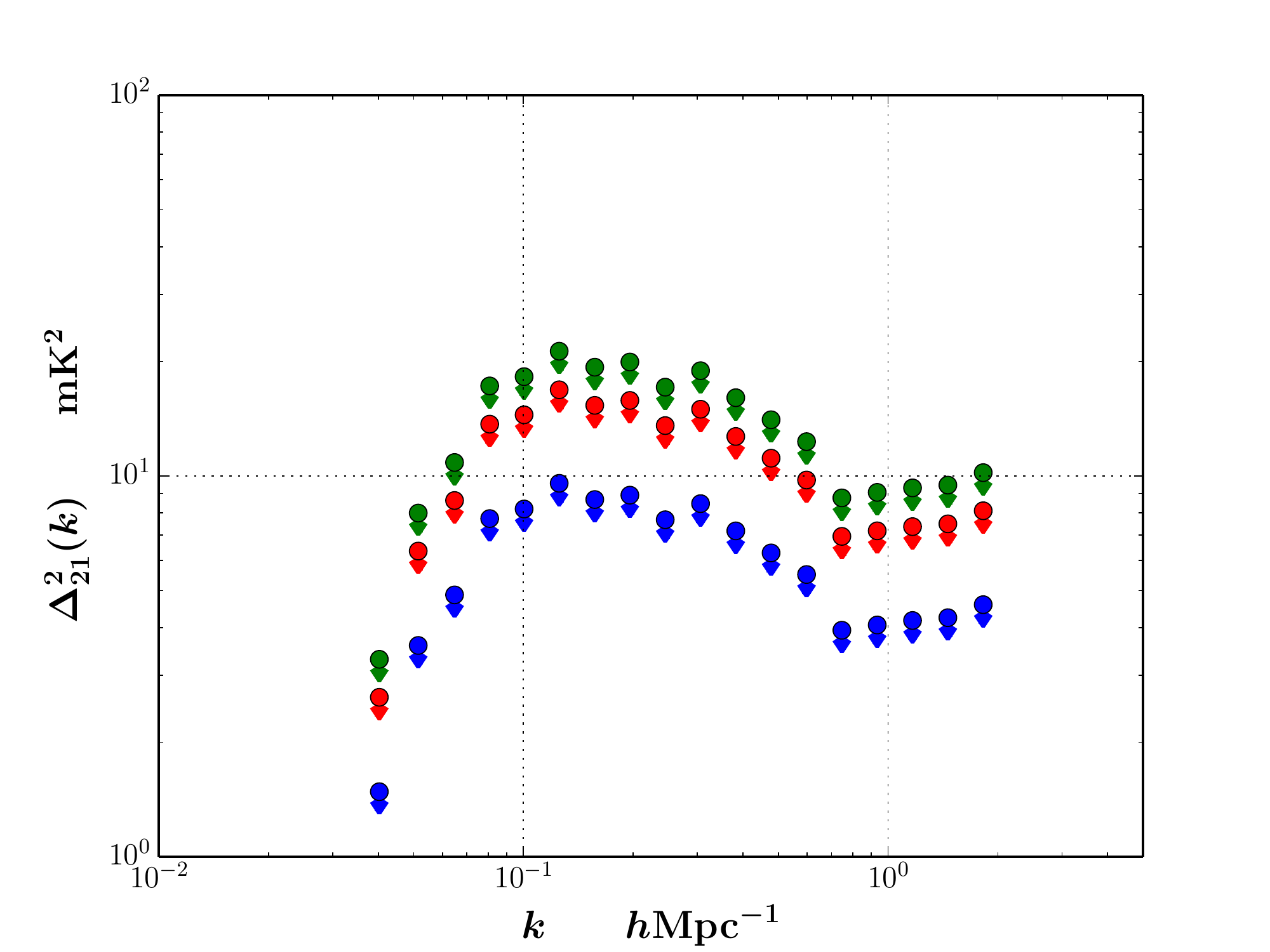}
\caption{Confidence upper limits on the amplitude of the $z = 5.9$ 21~cm power spectrum 
as a function of $k$.  Colors corresponds to 65 per cent confidence (blue), 95 per cent confidence (red),
and 99 per cent confidence (green).}
\label{fig:upperlimits}
\end{figure}

Figure \ref{fig:upperlimits} shows the main result of this paper: the
upper limits on the 21~cm power spectrum amplitude as a function of $k$.  We plot 65 per cent, 95 per cent,
and 99 per cent confidence limits in blue, red, and green, respectively.
In general, there is no significant variation in amplitude as a function of $k$
and most limits are within a factor of two of 10~${\rm mK}^2$.
While we do not plot the individual PDFs for each $k$ mode, we note that they are one-sided
Gaussian curves, with a peak at $\Delta^2(k) = 0\ {\rm mK}^2$.  The Gaussianity of the PDFs
results from treating the \cite{mcgreer_et_al_2015} limits as Gaussian.

Comparison of these results with typical \texttt{21cmFAST} outputs (e.g. Figure 4 of
\citealt{pober_et_al_2014}) show that the shape of the limits with $k$ is not surprising.
The knee feature due to the finite extent of ionized bubbles
exhibits itself as an absence of power on the largest scales, whereas the remaining $k$ modes
show a relatively flat power.
The peak in large-scale power around $k \sim 0.1\ h{\rm Mpc}^{-1}$ is set by models with highly biased 
sources (i.e. high $T_{\rm vir,min}$) and large values of $R_{\rm mfp}$.  This combination of parameters can 
result in large cosmic \HI patches left over in the final stages of reionization. 
On the other hand, the fall off in power towards high $k$ depends most strongly on the $R_{\rm mfp}$.  These constraints span a range in $k$ from approximately 0.05 to $2.0\ h{\rm Mpc}^{-1}$,
covering the region where first-generation 21~cm experiments will have maximal
sensitivity \citep{parsons_et_al_2012a,parsons_et_al_2012b,van_haarlem_et_al_2013,beardsley_et_al_2013,pober_et_al_2014}.

\section{Discussion}
\label{sec:discussion}

The principal implications of this result are simple: 21~cm cosmology experiments should not detect
power spectra above $\sim 10 - 20~{\rm mK}^2$ at $z=5.9$.  A detection of power above this level 
would be highly indicative of residual foregrounds or other systematic contamination of the power
spectrum.  This test should be used in conjunction with other measurements at higher redshifts.
That is to say, an experiment claiming a detection of cosmological power at $z > 6.0$ would be suspect
if it also detected power above this limit at $z = 5.9$.

To get a scale for this limit, we can compare it with both the current best limits on the 21~cm signal
and with expected power spectrum amplitudes for the reionization signal.
The current best limits on the 21~cm power spectrum come from \cite{ali_et_al_2015} at $z=8.4$,
constraining $\Delta^2(k) < 501.8\ {\rm mK}^2$ for $0.15\ h{\rm Mpc}^{-1} < k < 0.5\ h{\rm Mpc}^{-1}$. 
Limits from \cite{jacobs_et_al_2015} cover a range of other redshifts, including $z = 7.55$, but
are over a factor of four above the \cite{ali_et_al_2015} limits.
A significant increase in sensitivity over current limits is therefore needed to reach
the theoretical limits presented here; however, the steep spectral slope of sky noise
reduces the power spectrum noise level by a factor of nearly three from $z = 8.4$ to 5.9 (assuming
a spectrally flat receiver contribution of 100~K).
Most models, on the other hand, predict reionization power spectra with peak brightnesses of 
$\sim 10~{\rm mK}^2$ \citep{pober_et_al_2014,mesinger_et_al_2016},
on par with the limits presented here.
Therefore, with even a factor of two to three less integration time at $z=5.9$, 
any 21~cm experiment presenting a detection result at higher redshift
should have the sensitivity to compare with the results
presented here.

Estimates for the amount of integration time needed to make a first detection of the 21~cm signal
with current experiments range from several hundred to one-thousand hours 
\citep{morales_2005,mcquinn_et_al_2006,pen_et_al_2009,parsons_et_al_2012a,beardsley_et_al_2013,van_haarlem_et_al_2013,pober_et_al_2014}.
Therefore, even a factor of three less observing time for a null result
at $z = 5.9$ is a significant investment of resources.  For experiments with very wide instantaneous
bandwidths, $z = 5.9$ measurements come in effect ``for free."  Unfortunately, none of the first generation
experiments are quite able to achieve this.  The PAPER experiment has an instantaneous bandwidth
from 100 to 200 MHz ($z = 6.1$ to $z = 13.2$), with a somewhat smaller usable bandwidth due to 
analog filters and the difficulty of foreground removal near the end of the band
\citep{parsons_et_al_2010}.  The MWA is sensitive to frequencies between 80 and 300 MHz
($z = 3.7$ to $z = 16.8$), and so can easily target a redshift of 5.9; however, it has an
instantaneous bandwidth of only 30.72~MHz, and so would need to devote a separate observing
campaign for the $z = 5.9$ limit.  Depending on the Nyquist zone chosen,
LOFAR high band antennae have an instantaneous bandwidth covering either 
110 to 190 MHz ($z = 6.5$ to $z = 11.9$), 170 to 230
MHz ($z = 5.2$ to $z = 7.4$), or 210 - 250 MHz ($z = 5.2$ to $z = 5.8$) \citep{van_haarlem_et_al_2013}.
While either of the latter two modes would allow for a $z = 5.9$ test, it appears that most of LOFAR's
EoR observing has taken place in the first mode \citep{yatawatta_et_al_2013,jelic_et_al_2014}.
Next generation experiments like the Hydrogen Epoch of Reionization Array 
(HERA)\footnote{http://reionization.org} and the Square Kilometre Array 
(SKA)\footnote{http://skatelescope.org} are being designed with wider instantaneous bandwidths and
so can take advantage of this low redshift limit without devoting additional observing time.
HERA, in particular, is targeting an instantaneous bandwidth of $50-250$~MHz (DeBoer et al., \emph{in prep.});
taking into account the sensitivities for HERA calculated in DeBoer et al., \emph{in prep.}, HERA
should not only be able to use the upper limits here as a consistency check on 
its higher redshift measurements,
but should also be able to detect the presence of neutral gas in the IGM below $z =6$ with high significance.

Lastly, it is worth commenting on the uncertainties introduced in this calculation through our dependence
on \texttt{21cmFAST} models of reionization.  The three parameter ($\zeta,\ T_{\rm vir,min},\ R_{\rm mfp})$
model for reionization is certainly simplified, although as we note above,
it nevertheless serves to provide an effective basis set for 21~cm power spectra.
One significant approximation is that does not include any time or halo mass dependence 
in the parameters.  For this work, however, the lack of spatial fluctuations
in the ionizing photon mean free path can be considered the poorest approximation. 
The end stages of reionization
can be highly sensitive to the interplay between sources and sinks of ionizing photons, producing
non-trivial spatial inhomogeneities that are not captured by a uniform photon horizon
(e.g. \citealt{sobacchi_and_mesinger_2014}).
However, we note that more realistic models (e.g. \citealt{mesinger_et_al_2016}) predict significantly
less 21~cm power on moderate and large scales during the end of reionization.
In particular, the two reionization simulations in \cite{mesinger_et_al_2016} produce
21~cm power spectra with maximum values of $\sim2$ and $\sim7\ \mathrm{mK}^2$ (depending
on the bias of the galaxies responsible for the ionizing photons) when the universe
is 10 to 15 per cent neutral.
In this sense, our limits on the 21~cm signal are conservative;
including more physics in the simulations would further reduce the probability of a bright
21~cm signal at $z=5.9$.  

The EoR morphology of \cite{sobacchi_and_mesinger_2014} can be crudely
approximated by a small photon horizon.  
Lowering the $R_{\rm mfp}$ prior therefore should reproduce the general trend expected
from more realistic reionization simulations.
If we limit the range of $R_{\rm mfp}$ to less than 15~Mpc, our two and three
sigma upper limits at $k \lesssim\ 0.2 h{\rm Mpc}^{-1}$ decrease by a factor of $\sim 2$, 
with limits at higher $k$ effectively unchanged, again suggesting that our overall
limits are conservative.

We have also neglected to incorporate evolution effects across the redshift
axis of our \texttt{21cmFAST} simulations, i.e., the light cone effect 
\citep{datta_et_al_2012,datta_et_al_2014,la_plante_et_al_2014}.
Investigations into this effect find the largest scales of the 21~cm power spectrum to
be the most affected ($k \lesssim 0.1~h{\mathrm{Mpc}}^{-1}$), which are those most likely to be 
contaminated by foregrounds for the current and next generation of 21~cm experiments
(\cite{parsons_et_al_2012a,pober_et_al_2014}; Sims et al., in review).
More importantly, the general trend of the effect during the end
of reionization is reduce power on the larger scales by tens of per cent 
\citep{datta_et_al_2014}.  Therefore, in this regard, our upper limits on the 21~cm
signal are conservatively high.

\section{Conclusions}
\label{sec:conclusions}

We have presented an upper limit on the amplitude of the 21~cm power spectrum at a redshift of 5.9.  This
limit is derived from the model-independent limit on the neutral fraction from the dark pixel
analysis of quasar absorption spectra in \cite{mcgreer_et_al_2015}.  To convert a neutral
fraction limit to a 21~cm power spectrum one, we use a modified version of \texttt{21CMMC} to explore
the likelihood of 21~cm power spectra given the prior constraints.  The end result is a limit
of $\Delta^2(k) < 10$ to $20\ {\rm mK}^2$ at 99 per cent confidence, 
over a range of $k$ from roughly 0.5 to $2.0\ h{\rm Mpc}^{-1}$,
with the exact value depending on the $k$ mode in question.
This limit can be used as a null test for 21~cm experiments looking
to build confidence in any purported initial detection of the cosmological signal. 
While a non-detection of power at these redshifts is not definitive validation of
a higher redshift detection, any power
detected at $z = 5.9$ in excess of our limit is highly suggestive of residual 
foregrounds or systematic
errors that would likely also affect measurements at other redshifts.

\section*{Acknowledgements}

The authors would like to thank Steve Furlanetto and James Aguirre for helpful comments on a draft version
of this manuscript.
AM and BG acknowledge support from the European Research Council (ERC) under the European Union's Horizon 
2020 research and innovation program (Starting Grant No. 638809 --- AIDA --- PI: Mesinger).



\bibliographystyle{mnras}
\bibliography{/Users/jpober/Dropbox/bibfiles/masterbib}

\bsp	
\label{lastpage}
\end{document}